\documentclass[12pt,fleqn]{article}

\usepackage{amsmath,amssymb}
\usepackage{graphicx}
\usepackage[top=0.75in, bottom=0.75in, left=0.75in, right=0.75in, dvips]{geometry}

\begin{document}


\title{{\sf On the Standard Approach to Renormalization Group Improvement}}
\author{
F.A.\ Chishtie\thanks{Department of Applied Mathematics, The
University of Western Ontario, London, ON  N6A 5B7, Canada}, V.\
Elias\thanks{Department of Applied Mathematics, The University of
Western Ontario, London, ON  N6A 5B7, Canada}, R.B.\ Mann\thanks{
Department of Physics, University of Waterloo, Waterloo, ON N2L
3G1,  Canada}, D.G.C.\ McKeon\thanks{Department of Applied
Mathematics, The University of Western Ontario, London, ON N6A
5B7, Canada}, T.G.\ Steele\thanks{Department of Physics and
Engineering Physics, University of Saskatchewan, Saskatoon, SK,
S7N 5E2, Canada} }
\maketitle
\begin{center}
UWO-TH-06/15
\end{center}

\begin{abstract}
Two approaches to renormalization-group improvement are examined:
the substitution of the solutions of running couplings, masses and
fields into perturbatively computed quantities is compared with
the systematic sum of all the leading log ($LL$), next-to-leading
log ($NLL$) {\it etc.\ }contributions to radiatively corrected
processes, with $n$-loop expressions for the running quantities
being responsible for summing $N^{n}LL$ contributions. A detailed
comparison of these procedures is made in the context of the
effective potential $V$ in the 4-dimensional $O(4)$ massless
$\lambda \phi^{4}$ model, showing the distinction between these
procedures at two-loop order when considering the $NLL$
contributions to the effective potential $V$.
\end{abstract}

The process of renormalization necessarily introduces an arbitrary
renormalization scale $\mu$ into radiative processes in quantum
field theory. Requiring that a process
$\Gamma=\Gamma(\lambda,m,\phi,\mu)$ (with $\lambda$, $m$, $\phi$
being coupling, mass and external field) be independent of $\mu$
leads to the usual renormalization group (RG) equation.
\begin{equation}
\mu\frac{d\Gamma}{d\mu}=\left(\mu\frac{\partial}{\partial\mu}+\beta(\lambda)\frac{\partial}{\partial\lambda}
+\gamma_{m}(\lambda)m\frac{\partial}{\partial m}
+\gamma_{\phi}(\lambda)\phi\frac{\partial}{\partial\phi}
\right)\Gamma=0
\label{rg_eq}
\end{equation}
where 
\begin{gather}
\beta=\mu\frac{d\lambda}{d\mu}~,~\lambda(\mu_0)=\lambda_0,
\label{beta_def}\\
\gamma_m=\frac{\mu}{m}\frac{d m}{d\mu}~,~m(\mu_0)=m_0
\label{gammam_def}\\
\gamma_{\phi}=\frac{\mu}{\phi}\frac{d
\phi}{d\mu}~,~\phi(\mu_0)=\phi_0 \label{gamma_def}
\end{gather}

The usual procedure \cite{1} is to solve Eq.~(\ref{rg_eq}) by the
replacement
\begin{equation}
\Gamma(\lambda,m,\phi,\mu)\rightarrow\Gamma\left(\overline{\lambda}(t),\overline{m}(t),\overline{\phi}(t),\mu\right)
\end{equation}
by invoking the method of characteristics \cite{2}. Here
$\Gamma(\lambda,m,\phi,\mu)$ is a perturbatively computed
approximation to the exact expression for $\Gamma$, and
$\overline{\lambda}$, $\overline{m}$ and $\overline{\phi}$ are the
solutions of the running versions
Eqs.~(\ref{beta_def})--(\ref{gamma_def})
\begin{gather}
\frac{d\bar\lambda}{d t}=\beta\left(\bar\lambda(t)\right)~,~\bar
\lambda(0)=\lambda_0
\label{beta_run}\\
\frac{1}{\bar m}\frac{d \bar m}{d
t}=\gamma_m\left(\bar\lambda(t)\right)~,~\bar m(0)=m_0
\label{gammam_run}\\
\frac{1}{\bar\phi}\frac{d \bar\phi}{d
t}=\gamma_\phi\left(\bar\lambda(t)\right)~,~\bar \phi(0)=\phi_0
\label{gamma_run}
 \end{gather}
where $t=\log\frac{A^2}{\mu^2}$
 with $A^2$ typically (though not uniquely) specified by the energy scales of the
 process.

The incorporation of successively higher loop
contributions to the functions $\beta$, $\gamma_m$ and
$\gamma_{\phi}$
into
$\Gamma\left(\overline{\lambda}(t),\overline{m}(t),\overline{\phi}(t),\mu
\right)$
represents a standard method of RG improvement.
A detailed outline of this approach appears in Ch.~7 of Ref.~\cite{collins},
that has been applied, for
example, to the contribution of the effective potential $V$ in
\cite{3}. In this note we wish to examine this procedure in more
detail, and illustrate how its application compares with a systematic
sum of all the
leading log ($LL$), next-to-leading log ($NLL$) {\it etc.\
}contributions. In
particular, we look at the effective potential $V$ in the $O(4)$ massless
$\lambda {\phi}^{4}$ model\footnote{The $O(4)$ model is of particular interest as it represents the scalar field theory projection of the Standard Model.} beyond one loop order in the $NLL$
 approximation.

The tree-level potential for this model is
\begin{equation}
V_{tree} = \pi^2\phi^4 y
\label{V_tree}
\end{equation}
where $y \equiv\frac{\lambda}{4 \pi^2}$ denotes the scalar
couplant in the theory. At one loop order  we have
\cite{4},\footnote{We have dropped the subscript ``0" on $y$ and
$\phi$ for subsequent analysis.}
\begin{gather}
\frac{d\overline{y}(t)}{dt}=6\overline{y}(t)^2~,~\overline{y}(0)=y
\label{one_loop_y}\\
\frac{d\overline{\phi}(t)}{dt}=0~,~ \overline{\phi}(0)=\phi
\label{one_loop_phi}
\end{gather}
 Replacing $y$ and $\phi$ in Eq.\ (\ref{V_tree})
by the solutions to Eqs.\ (\ref{one_loop_y}--\ref{one_loop_phi})
\begin{gather}
\overline{y}(t)=y+6y^2t+O(t^2)
\label{y_oneloop}\\
\overline{\phi}(t)=\phi+O(t)~.
\label{phi_oneloop}
\end{gather}
results in
\begin{equation}
V_{tree} \rightarrow \pi^2\overline{\phi}(t)^4 \overline{y}(t)=
\pi^{2} \phi^{4} \left(y+6y^2t+\ldots\right)~.
\label{V_oneloop_improved}
\end{equation}
This expression agrees with the leading-logarithm ($LL$) terms in the
explicit calculation of the one-loop potential \cite{5,6,7,8},
\begin{equation}
V_{1loop}=\pi^2\phi^4 \left(y+6y^2t+T_{1,0}y^2+\ldots\right)
\label{V_oneloop}
\end{equation}
where $T_{1,0}$ denotes the non-logarithmic term whose value
depends on the renormalization scheme.\footnote{We have taken
$t=L=\frac{1}{2}\log\left(\frac{\phi^2}{\mu^2}\right)$.}
Thus the replacement of Eq.~(\ref{V_oneloop_improved}) does represent a systematic incorporation of the
$LL$ terms of the effective potential.

The general form of $V$ is given by
\begin{equation}
V=\pi^2\sum_{n=0}^{\infty}\sum_{m=0}^{n} y^{n+1}
T_{n,m}t^{m}\phi^4 =\pi^2\sum_{n=0}^{\infty} y^{n+1} S_{n}(yt)
\phi^4~,~T_{0,0}=1 \label{full_V}
\end{equation}
where $S_{n}(\xi)=\sum_{m=0}^{\infty}T_{n+m,m} \xi^{m}$. In Refs.~\cite{9,10} it is shown how Eq.~(\ref{rg_eq}) results in a series of ordinary
differential equations that can be solved in turn for $S_0(\xi)$,
$S_1(\xi),\ldots$ in terms of the functions $\beta(\lambda)$ and
$\gamma_{\phi}(\lambda)$, provided that the boundary values
$T_{n,0}=S_n(0)$ are known. The function $S_n(\lambda L)$ gives the $N^nLL$
contributions to $V$. In particular, $S_0=\frac{1}{1-6yt}$ and
$S_1=\frac{4T_{1,0}-6yt +13\log(1-6yt)}{4(1-6yt)^2}$, so that to
order $t$, we obtain
\begin{equation}
V_{LL}+V_{NLL}=\pi^2\phi^4(yS_0+y^2S_1)
\approx\pi^2\phi^4\left[y+T_{1,0}y^2+\left(6y^2+12T_{1,0}y^3-{21}y^3\right)t+O(t^2)\right]~,
\label{V_NLL}
\end{equation}
which is indeed the $NLL$ approximation to the
explicit  two-loop contributions to $V$
\begin{equation}
V_{1loop}+V_{2loop}=\pi^2\phi^4
\left[y+T_{1,0}y^2+{T_{2,0}y^3}+\left(6y^2+12T_{1,0}y^3-21y^3\right)t+O(t^2)\right]~.
\label{V_twoloop}
\end{equation}
The quantity $T_{2,0}$, which is explicitly computed by a two loop calculation, is the initial value for the $NNLL$ summation $S_2(yt)$.
As noted previously, we see from the $LL$ terms in the above expression that the replacement of
Eq.~(\ref{V_oneloop_improved}) does in fact generate the $LL$ expression
$V_{LL}=\pi^2\phi^4yS_0(yt)$. We now will check to see if a
similar replacement of $\lambda$ and $\phi$ by the two loop
contribution to $\overline{\lambda}(t)$ and $\overline{\phi}(t)$
can generate the $NLL$ results of either Eq.~(\ref{V_NLL}) or
(\ref{V_twoloop}). The relevant RG equations at two-loop level are
\cite{13},
\begin{gather}
\frac{d\overline{y}(t)}{dt}=6\overline{y}(t)^2-\frac{39}{2}\overline{y}(t)^3
\label{run_y_twoloop}\\
\frac{d\overline{\phi}(t)}{dt}=-\frac{3}{8}\overline{y}(t)^2\overline{\phi}(t)~,
\label{run_phi_twoloop}
\end{gather}
whose solutions are
\begin{gather}
\overline{y}(t)=y+\left(6y^2-\frac{39}{2}y^3\right)t+O(t^2)
\label{y_twoloop}\\
\overline{\phi}(t)=\phi-\frac{3}{8}y^2\phi t +O(t^2)~.
\label{phi_twoloop}
\end{gather}
Replacing $\phi$ and $y$ in $V_{1loop}$ of Eq.~(\ref{V_oneloop}) with the above results for
$\overline{\phi}(t)$ and $\overline{y}(t)$
results in
\begin{equation}
V_{1loop}\rightarrow\pi^2\phi^4\left[y+T_{1,0}y^2+\left(12y^2-21y^3+12T_{1,0}y^3-\frac{81}{2}T_{1,0}y^4
\right)t+O(t^2)\right]
\label{comp_one}
\end{equation}
which is distinct from both Eqs.~(\ref{V_NLL}) or
(\ref{V_twoloop}) at $LL$ level. Furthermore, substitution of equations
(\ref{y_twoloop}) and (\ref{phi_twoloop}) into the modified tree
potential $\widetilde{V}_{tree}=\pi^2\phi^4(y+T_{1,0}y^2)$ results
in
\begin{equation}
\widetilde{V}_{tree}\rightarrow\pi^2\overline{\phi}(t)^4\left[\overline{y}(t)+T_{1,0}\overline{y}(t)^2\right]
\approx\pi^2\phi^4\left[y+T_{1,0}y^2+\left(6y^2+12T_{1,0}y^3-21y^3-\frac{81}{2}T_{1,0}y^4\right)t
+O(t^2)\right]~.
\label{comp_two}
\end{equation}
This expression does agree with the $NLL$ terms in (\ref{V_NLL}) and (\ref{V_twoloop}), but it is not a systematic
$NLL$ expansion because is
contaminated with $N^2LL$ $y^4 t$ terms.

We thus see that simply replacing, in the one loop result, renormalized parameters by running parameters computed to two loop order,
does not reproduce the systematic $NLL$ approximation to  (\ref{V_twoloop})
obtained either from explicit calculation of the two loop graphs or
in (\ref{V_NLL})
from carefully solving the RG equation for the functions $S_0$ or $S_1$. In Section 7.6 of Ref.~\cite{collins} the recursion relations that fix the form of the functions $S_0$ and $S_1$ are obtained from the RG equation and a comparison with the substitution of ``running" RG group functions into perturbative results is suggested. It is this comparison that we have examined in detail above.
One should not conclude from our analysis that
the method of characteristics is inappropriate for
extracting information from the renormalization group about
physical processes. In Refs.~\cite{11,12} ways in which this
technique can be usefully applied are discussed.

It is worth briefly recapping the results of \cite{11,12} that are
pertinent to the discussion above. We first illustrate the method
of characteristics by considering the general equation
\begin{equation}
f(x,y)\frac{\partial A(x,y)}{\partial x}+g(x,y)\frac{\partial
A(x,y)}{\partial y}+h(x,y)A(x,y)=0 \label{char1}
\end{equation}
where $f$, $g$ and $h$ are known functions. If $A_0(x,y)$ is a
solution to Eq.\ (\ref{char1}), then it can be easily verified that
\begin{equation}
A(\bar{x}(\tau),\bar{y}(\tau),\tau)=A_0(\bar{x}(\tau),\bar{y(\tau)})\exp\left(\int_0^{\tau}h(\bar{x}(\tau'),\bar{y}(\tau'))d\tau'\right)
\label{char2}
\end{equation}
is also a solution, where $\bar{x}(\tau)$ and $\bar{y}(\tau)$ are
``characteristic functions" satisfying
\begin{gather}
\frac{d\bar{x}(\tau)}{d
\tau}=f(\bar{x}(\tau),y(\tau))~,~\bar{x}(0)=x \label{gen1}
\\
\frac{d\bar{y}(\tau)}{d \tau}=g(\bar{x}(\tau),y(\tau))~,~
\bar{y}(0)=y. \label{gen2}
\end{gather}

In fact, we can show from Eqs.~(\ref{char1},\ref{gen1},\ref{gen2})
that
\begin{equation}
\frac{d}{d \tau}A(\bar{x}(\tau),\bar{y}(\tau),\tau)=0,
\label{rgeq1}
\end{equation}
For example, if $f(x,y)=x$, $g(x,y)=y^2$, $h(x,y)=0$ and
$A_0(x,y)=xe^{\frac{1}{y}}$, then
\begin{gather}
\bar{x}(\tau)=xe^{\tau} \label{esol1}
\\
\bar{y}(\tau)=y(1-y\tau)^{-1} \label{esol2}
\end{gather}
and
\begin{equation}
A(\bar{x}(\tau),\bar{y}(\tau))=A_0(x,y)~.
\label{equiv1}
\end{equation}

However, if we only know the perturbative approximation to
$A_0(x,y)$
\begin{equation}
A_0^{(1)}(x,y)=x\left(1+\frac{1}{y}\right) \label{pert1}
\end{equation}
then
\begin{equation}
A_0^{(1)}(\bar{x}(\tau),\bar{y}(\tau))=xe^{\tau}\left(1+\frac{1-y\tau}{y}\right)
\label{pert2}
\end{equation}
which reduces to the exact solution $A_0(x,y)$ for one particular
value of $\tau$, namely $\frac{1}{y}$. Furthermore, if we
extrapolate from Eq.~(\ref{pert1}) assume the form of $A_0(x,y)$
to be
\begin{equation}
A_0(x,y)=x\sum_{n=0}^{\infty}\alpha_n\left(\frac{1}{y}\right)^{n}
\label{pert3}
\end{equation}
then the original differential equation [Eq.(\ref{char1})] leads to
the recursion relation
\begin{equation}
\alpha_{n}=(n+1)\alpha_{n+1} \label{rec}
\end{equation}
which shows that eq.(\ref{pert3}) is equivalent to
$xe^{\frac{1}{y}}$ if $\alpha_0=1$.

The points that arise in our discussion of this simple equation
have direct analogues when one analyzes the RG equation given in
Eq.~(\ref{rg_eq}). We first note that $\lambda$, $m$ and $\phi$ in
Eq.~(\ref{rg_eq}) are ``running" functions of $\mu$ with their
$\mu$ dependence dictated by
Eqs.~(\ref{beta_def}--\ref{gamma_run}). In applying the method of
characteristics to Eq.~(\ref{rg_eq}), we define ``characteristic"
functions $\bar{\lambda}(\tau)$, $\bar{m}(\tau)$,
$\bar{\phi}(\tau)$ by the equations
\begin{gather}
\beta(\bar{\lambda}(\tau))=\frac{d\bar{\lambda}(\tau)}{d\tau}~,~\bar{\lambda}(0)=\lambda(\mu)
\label{betac_def}\\
\gamma_m(\bar{\lambda}(\tau))=\frac{1}{\bar{m}(\tau)}\frac{d\bar{m}(\tau)}{d\tau}~,~\bar{m}(0)=m(\mu)
\label{gammamc_def}\\
\gamma_{\phi}(\bar{\lambda}(\tau))=\frac{1}{\bar{\phi}(\tau)}\frac{d\bar{\phi}(\tau)}{d\tau}~,~\bar{\phi}(0)=\phi(\mu).
\label{gammac_def}
\end{gather}
In Eqs.~(\ref{betac_def}--\ref{gammac_def}), we see that the
running functions act as boundary values for the characteristic
functions, while the form of
Eqs.~(\ref{beta_run}--\ref{gamma_run}) is identical to that of
Eqs.~(\ref{betac_def}--\ref{gammac_def}).

Just as substitution of the characteristic functions of
Eqs.~(\ref{esol1},\ref{esol2}) into the perturbative solution
(\ref{pert1}) reduces to the exact solution, so the substitution
of characteristic functions satisfying
Eqs.~(\ref{betac_def}--\ref{gammac_def}) into perturbatively
computed results for $V$ can reproduce exact solutions for $V$ in
terms of the functions $S_0,S_1,\ldots$ when the characteristic
parameter $\tau$ takes on a particular value. This is illustrated
in the context of massless scalar electrodynamics
in Section 7 of Ref.~\cite{12}.

We also note that determining an exact solution to
Eq.~(\ref{char1}) through substitution of Eq.~(\ref{pert3}) into
Eq.~(\ref{char1}) and thereby obtaining the recursion relation of
Eq.~(\ref{rec}) is directly analogous to the way in which the
functions $S_0$, $S_1$ etc. are generated in Refs.~\cite{9,10}.

\section*{Acknowledgements}
One of the authors, V. Elias, has recently passed away; the other
authors would particularly like to acknowledge their indebtedness
to him, both in this work and in many other collaborative efforts
with him. NSERC provided financial support.

\end{document}